# Steady-State Free Precession NMR in the Presence of Heteronuclear Couplings and Decoupling: More Than Meets the Eye

Sundaresan Jayanthi, Zuzana Osifová, Mark Shif, Adonis Lupulescu, and Lucio Frydman*

Department of Chemical and Biological Physics, Weizmann Institute of Science, Rehovot, Israel

**Abstract**

Fourier Transform (FT) has been a mainstay of analytical $^{13}$C/$^{15}$N NMR. On the other hand it has been shown that Steady-State Free Precession (SSFP) experiments which depart from this scheme can, under certain conditions, endow $^{13}$C/$^{15}$N small-molecule NMR with comparable sensitivity and resolution. SSFP is one of the earliest and most widely used NMR pulse sequences, yet its analyses have focused on isolated spin-1/2 ensembles such as water. The present study demonstrates that significant deviations from such isolated spin-1/2 behavior may occur when SSFP is applied in the presence of spin-spin couplings. Even in the simplest case supporting such couplings, a single $^{13}$C J-coupled to a $^1$H, departures from the isolated spin-1/2 behavior arise in the $^{13}$C SSFP response –both in the absence and in the presence of $^1$H spin decoupling. In the former case deviations are produced by the differential relaxation of antiphase two-spin terms generated by the pulse train; in the latter case, magnified interferences may arise between the SSFP pulses and the coherent perturbation arising upon $^1$H decoupling. Although both phenomena are also known in FT NMR, the spectral distortions that they will originate may be much larger in the SSFP case –particularly if interpulse delays in large flip-angle SSFP pulse trains "resonate" with the coupling perturbations. The origins of these effects are here analyzed for heteronuclear spin-1/2 systems and corroborated with $^{13}$C NMR SSFP experiments recorded under different conditions. Additional considerations aimed at magnifying or suppressing these effects, as well as extensions to more complex scenarios, are also briefly discussed.

# 1. Introduction

$^1$H-decoupled $^{13}$C and $^{15}$N FT NMR have been mainstays of analytical spectroscopy for over 50 years.[1-3] On the other hand, it has been recently shown that $^{13}$C and $^{15}$N NMR experiments based on the Steady-State Free Precession (SSFP) approach can, under certain conditions, provide equal sensitivity and resolution as FT-NMR counterparts.[4,5] The SSFP pulse sequence (Figure 1a) consists of a long train of equally spaced RF pulses with constant flip angles $\alpha$, applied at repetition times $TR \ll T_2, T_1$. SSFP's attractiveness derives from the fact that, if the transverse and longitudinal relaxation times $T_2$ and $T_1$ are similar and if a peak's position is *a priori* known, this experiment can generate in a continuous fashion a signal deriving from ca. 50% of the equilibrium magnetization $M_o$; no other NMR experiment can match this performance in terms of signal emitted per unit time. This was already noted in Carr's original 1958 paper,[6] which also derived how the transverse and longitudinal magnetization components will depend in SSFP on a peak's offset and on the flip angle $\alpha$ (Figures 1b-1d). These SSFP profiles are periodic, repeating themselves at regular $1/TR$ intervals. Notice how for large flip angles the transverse magnetization detected at the steady state is very small for zero offset, but close to maximal when the offset approaches $\pm 1/2TR$. Due to the experiment's features and constrains, SSFP is widely used when spectra are simple (i.e., peak positions known), when molecules are small and in solution (i.e., when $T_2 \approx T_1$), and when sensitivity/unit-time is at a premium –for instance, for field-locking an NMR magnet,[7] or for performing a cardiac MRI exam.[8]

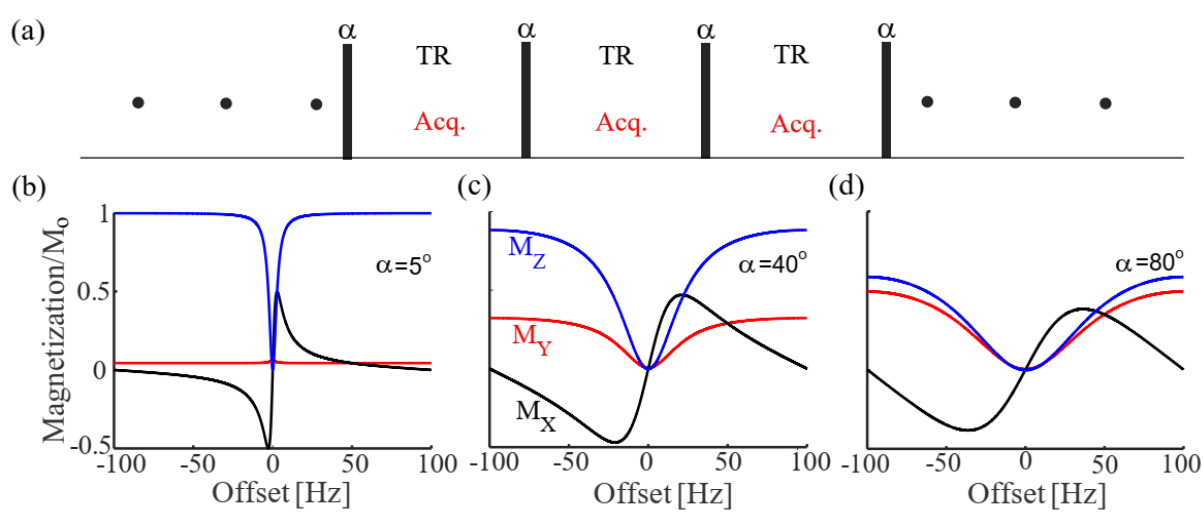

**Figure 1.** (a) SSFP pulse sequence consisting of a train of pulses with flip angle $\alpha$ and periods of free evolution spaced by repetition times $TR$, during which signals are acquired ("Acq."). The phase of consecutive pulses may be constant or incremented in a linear fashion, which is akin to changing the carrier's offset. (b-d) SSFP profiles arising for the steady-state magnetization components $M_X, M_Y, M_Z$

(black, red, blue respectively) for different angles as function of the offset of a 'single-spin' resonance with respect to carrier. The frequency profiles were calculated for TR = 5ms and $T_1 = T_2 = 1s$, and they represent the steady states measured immediately after the application of a pulse. Due to the periodicity of the sequence the magnetizations are periodic modulo $(1/TR)$ and hence are here plotted only over the $\pm 1/2TR$ intervals.

SSFP's features have been rederived, examined and exploited in numerous articles and studies; a contemporary literature search (Google Scholar®) gives over 20,000 references mentioning this experiment. However, to our knowledge, in all such studies but one[9] the focus of the SSFP experiment seems to have been on isolated single-spin systems such as the protons of water. During the aforementioned $^{13}$C SSFP high-resolution studies, however,[4,5] a series of anomalies were observed that showed these low-γ nuclei departing from the expectations arising for isolated single-spin ensembles. These departures arose in SSFP experiments carried out on simple natural abundance organic molecules subject to usual $^1$H broadband-decoupling protocols,[10-16] where each $^{13}$C site should have behaved as an isolated ensemble. But they also arose when SSFP sequences were executed in the absence of any $^1$H decoupling whatsoever, where the various components making up a given $^{13}$C multiplet could have been expected to behave as independent "single-spin" ensembles. The present study focuses on such observations, by revisiting the ca. 70-year-old SSFP experiment, for the simplest J-coupled system of a single $^{13}$C bound to a single $^1$H. It was found that significant deviations from the SSFP behavior expected for an isolated spin, may then indeed occur for the $^{13}$C signal –both in the presence and absence of $^1$H decoupling. In the presence of decoupling, these deviations will arise from interferences between the coherent perturbation imposed by the $^1$H irradiation, and the periodicity of the SSFP pulse train. In the absence of $^1$H decoupling, spontaneous $T_1$-driven fluctuations affecting the two-spin terms that arise over the course of the SSFP train, can also interfere with the expected $^{13}$C signal behavior. All these effects will depend on the timing, offset and the flip-angles used in the SSFP pulse train; this study explains these dependencies and exemplifies them for different coupling, decoupling and SSFP conditions on a simple $^{13}$C-$^1$H spin system. Ways to magnify, counter and exploit these effects, are also presented.

## 2. $^{13}$C SSFP NMR in the presence of a $^1$H J-coupling

We consider the simplest J-coupled system: a spin-1/2 ($^{13}$C) interacting with another spin-1/2 (a $^1$H) by a constant coupling ($J_{CH}$). The $^{13}$C will be assumed subject to an SSFP sequence where the pulses, separated by $TR$, are applied on-resonance at its chemical shift offset, and where the ensuing $^{13}$C signals are recorded in-between the pulses (Fig 1a). The only Hamiltonian to consider during the free evolution periods $TR$ of the SSFP sequence is thus

$$\mathcal{H} = 2\pi J_{CH} C_Z H_Z \qquad [1]$$

where spin operators have the usual meaning.[17,18] This Hamiltonian would normally generate a doublet, with "right" and "left" $^{13}$C peaks positioned at frequencies $\nu_C = \pm \frac{J_{IS}}{2}$ and having equal intensities. The corresponding observables for these peaks will be combinations of in-phase and anti-phase expectation values,

$$\langle P_\pm \rangle = \frac{\langle C_- \rangle \pm \langle 2C_- H_Z \rangle}{2} \qquad [2]$$

To derive analytical insight about how $\mathcal{H}$ will affect an SSFP experiment, also spin-relaxation needs to be considered. A simplified model was here assumed where the longitudinal and transverse relaxation times of the $^{13}$C and $^{1}$H were taken equal – $T_{1H} = T_{1C} = T_1$, $T_{2H} = T_{2C} = T_2$. Further, the two-spin zero-, single- and double-quantum coherences as well as the $C_Z H_Z$ state that may be generated by pulses, were also assumed to relax with the same $T_2$. While quite artificial, these assumptions could be relaxed without losing generality regarding the observed phenomena (*vide infra*).

With these assumptions one can compute the steady-states that will characterize a $^{13}$C SSFP experiment, based on the product between a propagator $U_{pulse}$ accounting for the action of the pulses (assumed here instantaneous), times the propagator $U_{free}$ accounting for evolution during the repetition period $TR$. These propagators were based on the 'improved' Liouvillian method which renders the master equation homogeneous.[19-21] $U_{free}$ incorporates the free precession phases $\phi = \pm \pi J_{CH} TR = \pm \theta_J$ arising from Eq. [1], while relaxation processes occurring during $TR$ are summarized by exponentials $E_1 = \exp(-R_1 \cdot TR), E_2 = \exp(-R_2 \cdot TR)$, where $R_{1,2} = 1/T_{1,2}$. The steady state conditions can then be found by extending Carr's magnetization-based analysis to Liouville-space density matrices $\rho$:

$$U_{pulse} U_{free} \rho = \rho \qquad [3]$$

that fulfil identity immediately after consecutive pulses. This model, further explained in Supplementary Information Section 1, then yields the steady state complex magnetization for arbitrary flip angles $\alpha$ for both the right-hand peak in the $^{13}$C multiplet:

$P_R = \langle P_+ \rangle =$

$$= \frac{i\,(-1+E_1)\sin\alpha\,(1+E_2^2+2E_2\cos^2\theta_J-2E_2\cos\alpha\sin^2\theta_J)}{2(-1+E_1E_2^3+E_2(-2+E_1-E_2+2E_1E_2)\cos^2\theta_J+(1+E_2)\cos\alpha\,(E_1-E_2^2+(-1+E_1)E_2\cos2\theta_J)+E_2(-E_1+E_2)\cos^2\alpha\sin^2\theta_J)} \quad [4]$$

and for the left-hand peak

$P_L = \langle P_- \rangle =$

$$= \frac{(-1+E_1)\sin\alpha\,(1+E_2)(i\cos\theta+\sin\theta)((1+E_2)\cos\theta-i(-1+E_2)\sin\theta)}{2(-1+E_1E_2^3+E_2(-2+E_1-E_2+2E_1E_2)\cos^2\theta_J+(1+E_2)\cos\alpha\,(E_1-E_2^2+(-1+E_1)E_2\cos2\theta_J)+E_2(-E_1+E_2)\cos^2\alpha\sin^2\theta_J)} \quad [5]$$

where $M_o$ was assumed unity.

While Eqs. [4] and [5] do not provide immediate insight about the SSFP behaviour of the $^{13}$C multiplet, we consider two particular cases that help illustrate the difference between SSFP on isolated spins vs SSFP in J-coupled spins:

Case (A). Consider a situation where $TR$ and offsets are chosen such that both peaks of the $^{13}$C J-split doublet, will fall at maxima of the SSFP profiles (Fig. 2a). This will require that $\theta_J = \pi J_{CH} TR$ equals an even multiple of $\frac{\pi}{2}$: $\{\theta_J = k\pi\}_{k=1,2,3...}$; i.e. TR $= k/J_{CH}$. Setting the "right" peak of the $^{13}$C spin doublet to an offset of $-1/2TR$ can then lead to a situation like that shown in Fig. 2a; the steady state magnetizations of both left and the right legs of the $^{13}$C doublet will then have the same normalized intensities:

$$P_R(A) = P_L(A) = \frac{i\,(-1+E_1)\sin\alpha}{2(-1+E_1E_2+(E_1-E_2)\cos\alpha)}. \quad [6]$$

Case (B). Consider now a situation where $TR$ is chosen in such way that the right-hand peak of the $^{13}$C J-split multiplet is again at the same SSFP maximum, but its left peak counterpart is at a minimum of the SSFP profile (Fig. 2b). For this to happen $\theta_J$ needs to be an odd multiple of $\frac{\pi}{2}$: $\{\theta_J = \frac{(2k+1)\pi}{2}\}_{k=0,1,2,3...}$; in other words, TR $= (2k+1)/(2J_{IS})$. In this instance the steady state magnetizations for the right and left hand-sides of the $^{13}$C doublet will differ, and be given by

$$P_R(B) = \frac{i\,(-1+E_1)\sin\alpha\,(1+E_2^2-2E_2\cos\alpha)}{2(-1+E_1E_2^3+(1-E_2^2)\cos\alpha\,(E_1+E_2)+E_2(-E_1+E_2)\cos^2\alpha))} \quad [7]$$

$$P_L(B) = \frac{i\,(-1+E_1)\sin\alpha\,(1-E_2^2)}{2(-1+E_1E_2^3+(1-E_2^2)\cos\alpha\,(E_1+E_2)+E_2(-E_1+E_2)\cos^2\alpha))}. \quad [8]$$

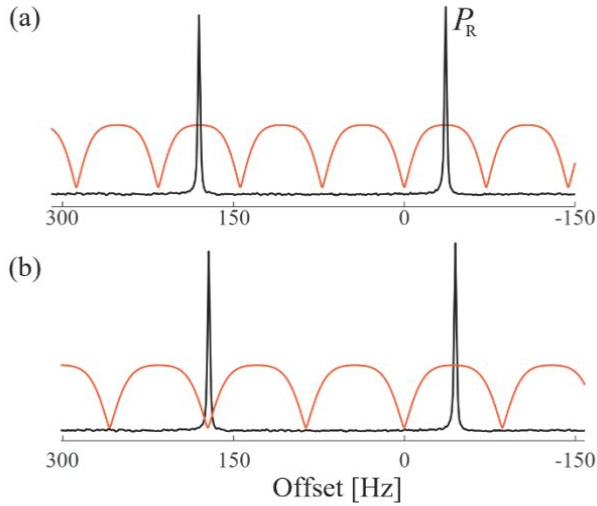

**Figure 2.** Two situations highlighting the influence of a J-coupling on the intensity arising from SSFP for a given multiplet. (a) Case where the J-doublet (in this case, for the $^{13}C$ chloroform peak) is chosen such that the frequencies of both components coincide with maxima of an isolated-spin SSFP profile (whose transverse magnetization is shown overlaid in red for a TR = 13.9 ms). (b) Same as in (a) except that TR = 11.6 ms, so that the frequency of left-hand component coincides with a minimum of SSFP profile while the right-hand side of the doublet remains at an SSFP maximum.

To highlight the differential effects that J-couplings have on SSFP, we focus on the magnetization $P_R$ –always chosen to be at the same SSFP maximum– for each of these two cases. Further, we consider for simplicity a situation where $\alpha \approx 90°$ such that $cos\alpha \sim 0$ and $sin\alpha \sim 1$, and use the $TR \ll T_1, T_2$ assumption to retain only the linear terms in $E_1$'s and $E_2$'s Taylor expansions. It can then be shown that although the isolated-spin SSFP formalism predicts that the $P_R$s should be identical in the two cases, their values will be differentially affected by the J-coupling. Indeed, for the above-mentioned approximations, their ratio can be written on the basis of relaxation rates as

$$\frac{P_R(B)}{P_R(A)} = \frac{2(R_1 + R_2)}{R_1 + 3R_2} \qquad [9]$$

When $T_1 = T_2$, $R_1 = R_2$ and the SSFP $P_R$ will be the same for both cases; however, for $T_1 > T_2$, $P_R(B)/P_R(A) < 1$ and thereby the intensity of this component will no longer be the same for cases (A) and (B): $P_R$'s intensity upon SSFP has been affected by the J-coupling. This is further analyzed in Figure 3a for a series of predictions deriving from Eqs. [4] and [5]. Figure 3b extends these expectations based on Liouville-space simulations performed using the Spinach package (and the $'r_1, r_2'$ relaxation flag for modelling the anti-phase, zero-quantum, double-quantum and spin-order relaxation rates);[22,23] this treats a J-coupled C-H spin model incorporating molecular tumbling as well as a third C-H---H spin. Calculating the dependence of the SSFP responses with respect to TR for various relaxation rates fulfilling $R_{1H} \neq R_{1C}, R_{2H} \neq R_{2C}$ shows results that are not significantly different from the highly simplified scenario leading to Figure 3a. Supplementary Information Section 2 provides further details on these calculations.

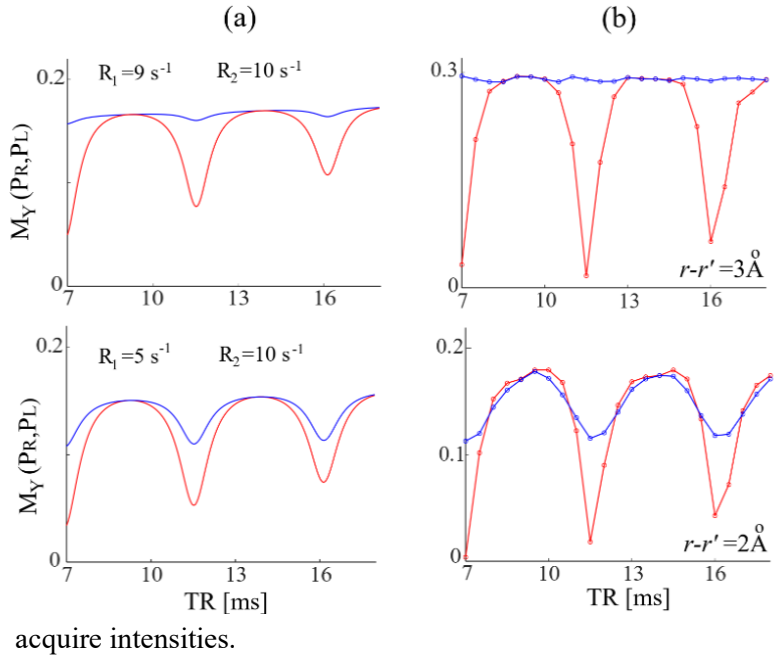

**Figure 3.** Variation of $P_R$ (blue) and $P_L$ (red) components of a $^{13}C$ doublet, as a function of $TR$. (a) Predictions arising from Eqs. [4] and [5], for the relaxation rates shown in the panels. (b) Spinach simulations showing the variation of $P_R$ (blue) and $P_L$ (red) as a function of $TR$ for a three-spin model system $I' \ldots I - S$; where $I$ and $S$ ($^1H$ and $^{13}C$) are J coupled. The distance $r(I',I)$ was taken as 3 Å on the top and 2 Å in the bottom panels, bringing in effect an increase in the relaxation rate $R_{1I}$. All calculations were performed assuming $\alpha = 40°$; these simulations in (b) are shown scaled with respect to 1D 90° pulse-acquire intensities.

Figure 4 corroborates these effects experimentally, with SSFP measurements as a function of different $T_{1H}$ values for a $^{13}C$ chloroform/$d_6$-DMSO sample, doped with increasing concentrations of Cr(acac)$_3$. This relaxation agent is known to shorten $T_1$s but have minor effects on $T_2$s;[24] for the case of chloroform solutions we also find that these changes are much more significant for the $^1H$ than for the $^{13}C$. As in the calculations of Figs. 2 and 3, experiments were here done by keeping the right-hand side of the doublet always at a maximum of an isolated-spin SSFP profile, and changing TR so that the left-hand peak in the doublet would experience subsequent maxima and minima of such profile. Notice how the addition of the $T_1^H$ relaxation agent will, by changing the $T_{1H}/T_{2H}$ ratio, enable the left-hand leg of the multiplet to significantly affect the right-hand counterpart, whenever the former goes through a trough of the SSFP profile. This –despite the latter being always placed at the same offset vis-à-vis the SSFP profile.

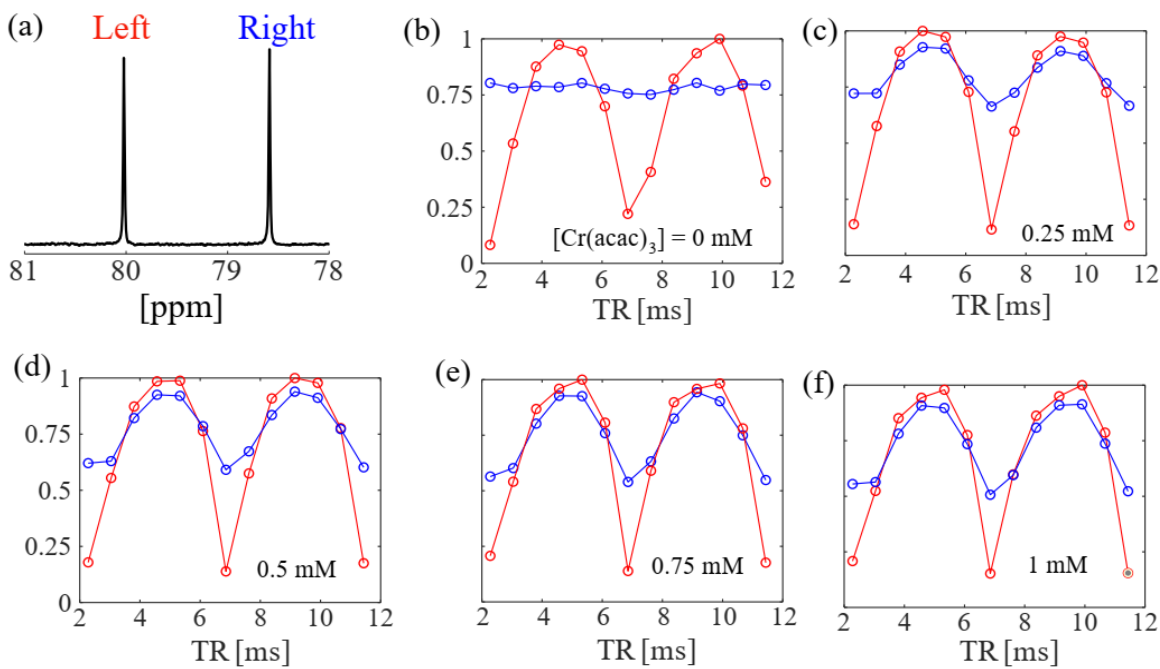

**Figure 4.** (a) $^1$H-coupled $^{13}$C chloroform spectrum labelling the right and left legs of its J-multiplet. In the experiments shown in this figure, the right-hand component was always kept at an SSFP profile maximum by placing the carrier on-resonance with it and subjecting pulses to a ±x phase alternation (Supporting Figure S2; this is equivalent to placing the carrier at a 1/2TR offset from it). (b-f) SSFP intensities of the two doublet components as function of $TR$ for different Cr(acac)$_3$ concentrations. The intensities of left (red) and right (blue) components of the doublet were normalized with respect to the global maximum in each plot. All experiments were performed with α = 40°. These points arose from high resolution spectra acquired with a conventional (~1.2 s) acquisition time and FT analysis, after the steady state was attained; this was done to obtain a clear differentiation between the chloroform doublet peaks, and between the chloroform and DMSO $^{13}$C signals. See Supplementary Information Section 3 for further details

Figures 2-4 demonstrate that the intensity of any one of the J-multiplet $^{13}$C components (e.g., $P_R$) will depend on the conditions created by the SSFP at the position of the other components (e.g., $P_L$). This can be understood from the different coherences that will be present at the time of the pulses applied during the SSFP experiment. Since in case (A) $\theta_J = k\pi$, no anti-phase coherences will be present right at the application of the pulses: the timing is chosen such that $C - H_Z$ coherences are identically zero at the pulses, and thus they not affect the spin dynamics. In case (B) by contrast $\theta_J = (2k + 1)\pi/2$; anti-phase terms will then be maximal at the position of the pulses, and hence affect the outcome of the sequence. This explains the reduction in the SSFP signal in case (B), as then the relative relaxation rates of $\{C_i\}_{i=x,y,z}$ and $\{C_i H_Z\}_{i=x,y,z}$ –mostly the faster decay of the latter– will influence the $^{13}$C component's intensity. These effects were here derived for a two-spin heteronuclear system, but similar effects can be expected when $^{13}$C or $^{15}$N are bonded to multiple protons.

# 3. $^{13}$C SSFP in the presence of $^1$H decoupling

While the preceding Section showed that J-couplings may influence SSFP peak intensities depending on the sequence's timing and on spin relaxation conditions, one could assume that J$_{CH}$-related effects would vanish in the presence of heteronuclear decoupling. Furthermore, decoupling is generally an integral constituency of $^{13}$C or $^{15}$N NMR acquisitions. While originally proposed based on a continuous wave (CW) irradiation of the heteronucleus to be decoupled,[25-27] experiments rely nowadays on composite pulses and/or adiabatic sweeps[10-16] that are repeated as supercycles of a basic decoupling unit for alleviating unwanted resonance offset effects and/or radiofrequency (RF) inhomogeneities. A common feature among all these schemes is the potential appearance of decoupling sidebands, reflecting the cyclicity imposed by the RF $^1$H irradiation on the spins' evolution. For most decoupling sequences operating under conventional conditions the amplitude of these sidebands is extremely low, and their presence barely noticeable. Nevertheless, these sidebands can be troublesome for certain types of experiments relying on long saturation pulses[28,29]. It turns out that under certain conditions heteronuclear decoupling sidebands may also unduly affect intensities in SSFP experiments, which as mentioned also rely on long ($\gg T_1$) pulse trains and will also have steep excitation and saturation offset dependencies –including a periodic nulling of the $M_z$ component (Figure 1).

To investigate these effects further we focus on what is arguably the simplest case where decoupling sidebands arise: the continuous on-resonance RF irradiation of a $^1$H, in an isolated $^{13}$C-$^1$H system. Our focus will be on the $^{13}$C, for which the relevant Hamiltonian is now

$$\mathcal{H}(t) = \omega_C C_Z + 2\pi J_{CH} H_Z C_Z + \omega_1 H_X \qquad [10]$$

where $\omega_C$ is the $^{13}$C offset, $\omega_1 = 2\pi\nu_1$ is the time-independent RF decoupling amplitude, and all remaining terms have their usual meaning. This Hamiltonian admits an exact solution of its Schrödinger equation,[1] leading to first-order sidebands at $\omega_C \pm \sqrt{\omega_1^2 + (\pi J_{CH})^2}$. An experimental observation of this prediction is shown Figure 5 for chloroform's isolated $^{13}$C-$^1$H pair, overlaid with an ideal single-spin SSFP profile like that expected for the $^{13}$C in the absence of couplings. This simple picture can help understand why decoupling may affect the SSFP $^{13}$C main peak intensity: if offset and repetition time $TR$ are chosen such that the decoupling sidebands coincide with the dips imparted by SSFP on $C_z$, the centerband's intensity may drop –even if the maximum is placed at a maximum of the SSFP profile, and even if decoupling sidebands amount to only a small percent of the main peak's intensity. By contrast, SSFP conditions that do not affect the $^{13}$C magnetization at the sidebands' positions, should leave the centerband free from amplitude losses.

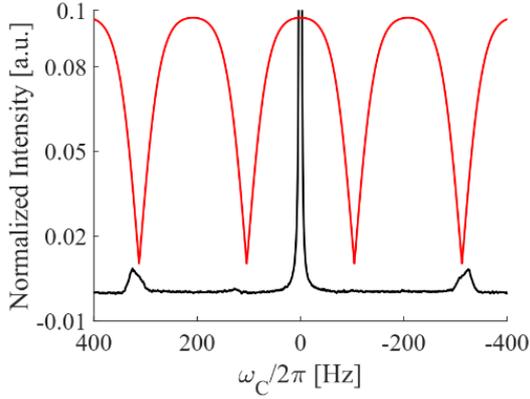

**Figure 5.** SSFP profile for an isolated spin ($M_Y$, in red) expected for $TR = 5$ ms and flip angle $\alpha = 80°$. Overlaid (in black) is CHCl$_3$'s $^{13}$C NMR spectrum obtained with proton CW decoupling ($\nu_1 = 300$ Hz, d$_6$-DMSO as solvent). The height of the spectrum was scaled so that the weak decoupling sidebands are visible, in a scale where the centerband's intensity was 1. The decoupling sidebands appear somewhat broadener than the centerband due to RF inhomogeneity effects.

While the Schrödinger equation for Eq. [10] can be readily solved, the generation by the decoupling/SSFP combination of all possible coherences for the two-spin system under consideration, coupled to the complexities of each term's longitudinal and transverse spin relaxation effects, leads to propagator expressions for which we were unable to derive generic steady state $^{13}$C solutions. In their stead, we consider again two cases akin to those treated in Section 2, together with some simplifying assumptions (on-resonance $^1$H decoupling, on-resonance $^{13}$C SSFP, and $R_{1H} = R_{1C} = R_1$; $R_{2H} = R_{2C} = R_2$) that also in this case were found to be secondary physics-wise. These were

Case (A). $TR = k/\nu_e$, where $|\nu_e| = \sqrt{\nu_1^2 + \frac{J_{IS}^2}{4}}$ is the decoupling sideband frequency and $k = 1,2,3...$ This corresponds to conditions where both the centerband and the decoupling sidebands coincide with maxima of the SSFP profile. Expression for the $^{13}$C SSFP centerband signals can then be derived; for $k = 1$ these will lead to transverse steady-state components

$$C_X(A) = 0; \quad C_Y(A) = \frac{\sin\alpha}{1 + e^{\frac{-R_2}{\nu_e}}} \qquad [11]$$

Other $k$ values lead to similar conditions, as described in Supplementary Information Section 4.

Case (B). Now $TR = (2k + 1)/2\nu_e$ and $k = 0,1,2...$ This corresponds to conditions where the decoupling sideband frequency coincides with a dip of the SSFP $C_z$ profile. The respective analytical expression for $k = 0$ is then

$$C_X(B) = 0; \quad C_Y(B) = \frac{(1 - e^{\frac{-R_2}{\nu_e}})\sin\alpha}{2(1 + e^{\frac{-2R_2}{\nu_e}} - 2e^{\frac{-R_2}{\nu_e}}\cos\alpha)} \qquad [12]$$

As in nearly all instances the $^{13}$C transverse relaxation rate will be much smaller than the decoupling RF frequency, it is possible to assume that $R_2 \ll \nu_e$ and $e^{\frac{-R_2}{\nu_e}} \approx 1$. $C_Y(A)$ will thus be much larger than $C_Y(B)$ for all $\nu_1$ RF fields of relevance, as illustrated in Figure 6 experimentally for a variety of decoupling conditions.

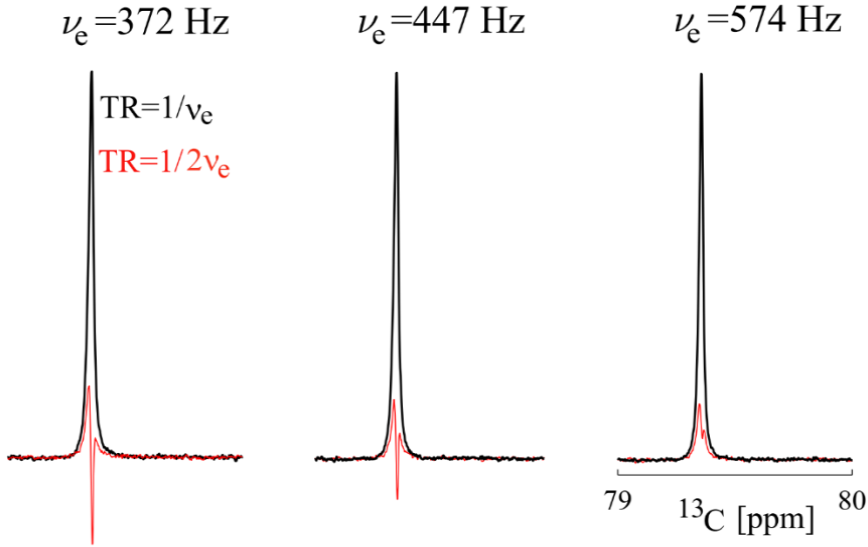

**Figure 6.** SSFP-defined $^{13}$C NMR high-resolution spectrum of CHCl$_3$ in d$_6$-DMSO obtained with proton CW decoupling, recorded for the conditions indicated in Eqs. [11] (black) and [12] (red) for three different $^1$H decoupling RF powers. Experiments were performed with $\alpha = 40°$ using the sequence in Supporting Figure S2b. This yielded high resolution spectra acquired with a conventional acquisition and FT analysis after the steady state was attained; this was done to obtain a clear differentiation between the chloroform and DMSO signals. Further details in Supplementary Information Section 3.

Eqs. [11], [12] can be extrapolated to other $k$-values and arbitrary RF fields, for both cases (A) and (B). For $TR = k/\nu_e$ we find that

$$C_X(A) = 0; \quad C_Y(A) = \frac{\sin\alpha}{1 + e^{\frac{-kR_2}{\nu_e}}} \qquad [13]$$

While for $TR = (2k+1)/2\nu_e$

$$C_X(B) = 0; \quad C_Y(B) = \frac{(1 - e^{\frac{-R_2(2k+1)}{\nu_e}})\sin\alpha}{2(1 + e^{\frac{-2R_2(2k+1)}{\nu_e}} - 2e^{\frac{-R_2(2k+1)}{\nu_e}}\cos\alpha)} \qquad [14]$$

The dependence of Eqs. [13], [14] on $k$ is shown in Supplementary Information Section S4, which also presents experiments corroborating the dependence of the signals on $TR$ and $k$, for different $\nu_e$ values.

Figure 7 explores these effects further, by presenting how chloroform's $^{13}$C NMR center peak will oscillate as a function of $\nu_1$ for fixed values of $TR$; these experimental data (red points) are compared against numerical simulations (black curves). Different panels also show how these centerband's dependencies will vary when SSFP experiments are performed for different $TR$ and different flip-angles $\alpha$. Notice how, as $\nu_1 \approx \nu_e$, the centerband intensity will go through peaks and troughs as the RF transverses from case (A) to case (B) and back to case (A). For low $\nu_1$ values these dips may become much stronger than the decoupling sidebands intensities vis-à-vis the centerband in the conventional FT spectrum, reflecting the aforementioned "magnification" effects. As $\nu_1$ increases however, the overall depth of these oscillations becomes progressively smaller; notice also the marked dependence of these effects on the flip angle, which follows from the wider SSFP saturation "dips" imparted on the z-magnetization with increasing flip angles (Figure 1). All this is as predicted by Eq. [14]; see Supplementary Information Section 5 for further details on this dependence.

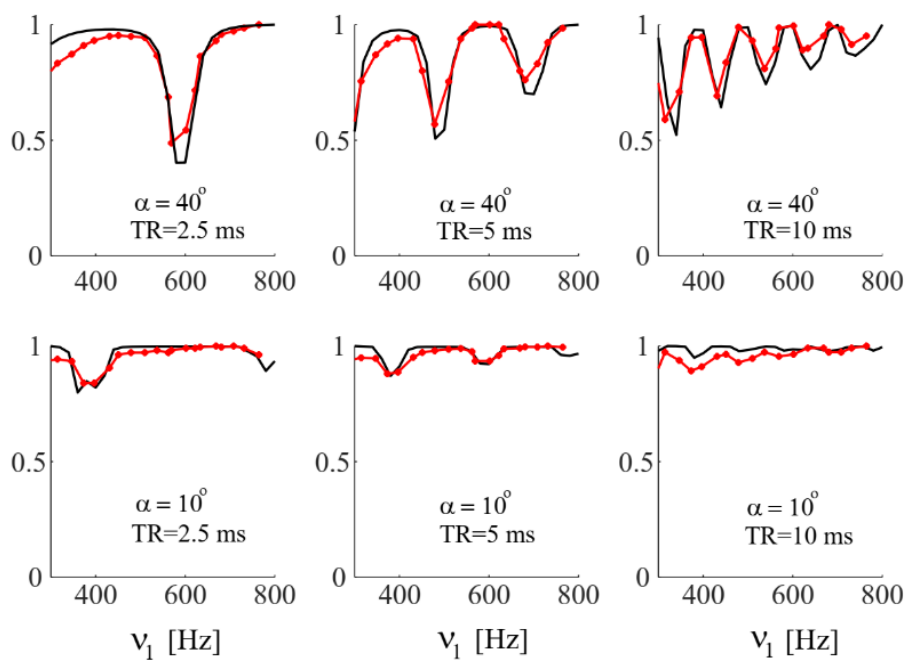

**Figure 7.** $^{13}$C SSFP signal of CHCl$_3$ in d$_6$-DMSO, recorded as function of the CW $^1$H decoupling amplitude, for different flip angles and repetition times. All curves are shown normalized with respect to their respective maximum values. Dots are experimental data, with red lines meant to serve as visual guides. The black curves represent outputs of Spinach simulations which incorporate an RF inhomogeneity of 10%. The $r_1, r_2'$ relaxation model was considered with $R_{1H} = 0.7\ s^{-1}$, $R_{1C} = 0.6\ s^{-1}$, $R_{2H} = R_{2C} = 1.5\ s^{-1}$. Experimental details are akin as in Figure 6.

While CW decoupling provides an amenable test case for comparing theory and experiments, modern $^{13}$C/$^{15}$N NMR is normally carried out utilizing more sophisticated $^1$H irradiation schemes. Figure 8 illustrates chloroform's $^{13}$C SSFP response when subject to WALTZ-16 decoupling,[30] for a series of $TR$ and $\alpha$ values. The $^{13}$C centerband was in these experiments placed on-resonance and excited with an x/-x phase alternating scheme, leading to maximal emission for the chosen flip-angle $\alpha = 40°$. WALTZ-16 is a highly-supercycled approach leading to numerous, usually negligible sidebands surrounding the main $^{13}$C decoupled peak. Notice by contrast how, whenever $1/TR$ hits one of those sidebands, the SSFP $^{13}$C response of chloroform shows a severe dip in intensity –akin to that described above for the CW case. The simplest way to avoid these signal losses seems to be to increase the $^1$H RF decoupling field; this in turn will decrease the length of the WALTZ-16 supercycles, leading to the nearly steady behaviours when the supercycle periods exceed ca. 500 Hz: the $^{13}$C SSFP emission becomes then maximal regardless of TR. Alternatively, reducing the flip angle also has a dramatic influence on the extent of these effects.

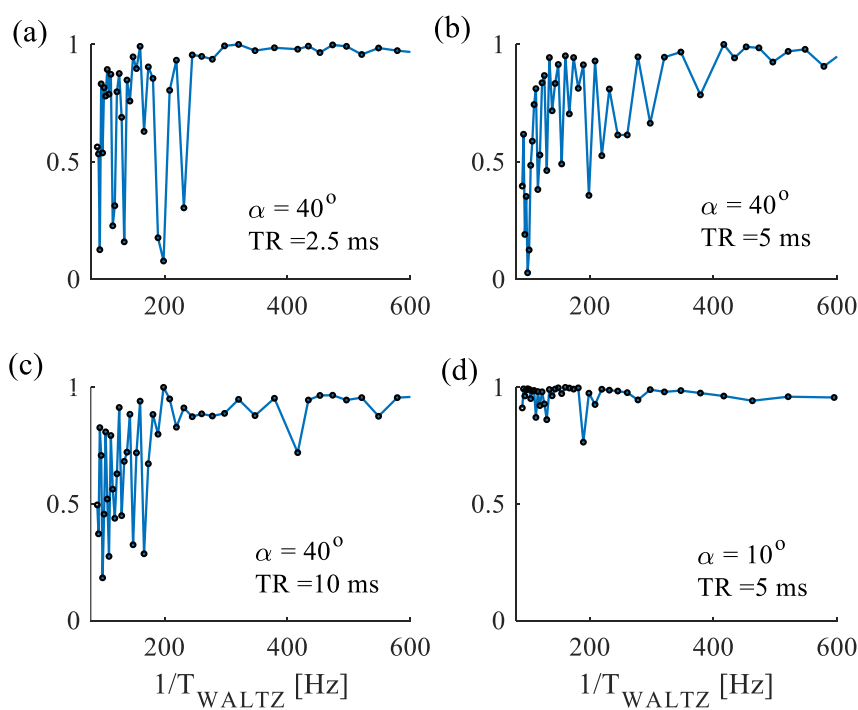

**Figure 8.** $^{13}$C SSFP signal of CHCl$_3$ in DMSO, as function of the inverse of WALTZ-16 cycle duration, $T_{WALTZ}$, for different flip angles and repetition times. The blue line connecting the black data is for guideline. All curves are normalized to their respective maximum values. Experiments were carried out and processed as described in Figures 6 and 7.

Finally, it is interesting to discuss the build-up and steady state values that other $I-S$ coherences may reach in the simple two-spin system. As shown in Supporting Information Section 6, only zero-quantum $ZQ_Y = (I_+S_- - I_-S_+)/2i = I_YS_X - I_XS_Y$ and double-quantum $DQ_Y = (I_+S_+ - I_-S_-)/2i = I_YS_X + I_XS_Y$ states are created with significant amplitudes, by the CW decoupling and SSFP combination. By contrast, although S-spin anti-phase coherences are non-zero over the $k \cdot TR < t < (k+1) \cdot TR$ intervals, they become immediately zero before or after any pulse of the SSFP sequence for both cases (a) and (b).

## 4. Discussion and Conclusions

Despite SSFP being one of the earliest and most widely used multi-pulse experiments, it appears that not all its aspects have yet been explored. The SSFP behaviour for an ensemble of isolated spins has been widely treated and exploited in both NMR and MRI, with analytical predictions that have been amply confirmed by both numerical calculations and experimental measurements. However, when executing $^{13}$C SSFP experiments as a function of finely incremented offsets as demanded by the recently PI-SSFP NMR proposal,[4-5] clear deviations from these expectations were seen in the signal intensities. These arose both when relying on $^1$H decoupling, but also when observing fully coupled spectra. Eventually, these deviations were traced to the new phenomena introduced in this study. When in the presence of decoupling, the PI-SSFP incrementation of the SSFP profile through a series of closely-spaced offsets eventually hit a WALTZ/GARP-derived sideband condition, leading to unexpected drops in a centerband's intensity. This phenomenon is akin to interferences we have observed between SSFP and magic-angle-spinning (MAS)[31] –although in a way it is easier to treat as it does not need to deal with the heterogeneities arising in the case of powdered samples. In the same way as for the MAS case, the phenomenon will decrease in intensity and eventually become negligible for high enough "rates of rotation" –happening in spin-space for the case of decoupling, and in actual space for MAS. Alternatively, the option of relying on randomized decoupling cycles also arises in the case of decoupling.

The second class of phenomena here analyzed arose even in the absence of decoupling, and concerned the intensity modulations observed under SSFP from various legs of a J-split multiplet, as offsets and TRs are changed. These intensities variations were found to depend on whether other legs in the multiplet experienced or not partial saturation, as driven by SSFP's periodic "dips". The fact that $1/TR$ in most SSFP experiments tends to be on the order of 100 Hz, makes this a commonly observed feature in the spectroscopy of coupled low-γ nuclei – particularly when targeting organic compounds, where J$_{CH}$ and J$_{NH}$ one-bond couplings are of the same magnitude. The strength of the ensuing effect will depend on the longitudinal relaxation times of the $^1$H, as relaxation processes are the means by which the various J-split

$^{13}$C/$^{15}$N components "communicate" with one another. As was also the case for the sideband-derived intensity anomalies arising in the presence of $^1$H decoupling, these effects will be highly dependent on the flip angles used during the SSFP scan. In both cases interference phenomena will be smaller for experiments relying on small SSFP flip angles; however, if relying on the larger flip angles that maximize SSFP's sensitivity for small-molecule heteronuclear NMR cases, these phenomena become easily observable.

The present study focused on the simplest spin-spin coupling case, involving a pair of heteronuclear spin-1/2s linked by a constant J. The situation is not expected to change significantly if several additional equivalent protons or if other kinds of spin-1/2 are added into the picture; however, we found that additional complexities (and opportunities) arise when subjecting homonuclear J-coupled systems to SSFP trains. Complexities that are not treated here will also occur if couplings are time-dependent –either because of MAS, or because of chemical exchange. These and other hitherto unexplored effects that could have an impact in current and future analytical uses of SSFP NMR,[32-34] are currently under investigation.

**Acknowledgments:** Support from the Israel Science Foundation (grant 1874/22), the Minerva Foundation, the EU program "SteadyNMR", and the Perlman Family Foundation, are acknowledged. SJ acknowledges Weizmann Institute for a sabbatical fellowship; ZO thanks the Azrieli Foundation for a postdoctoral fellowship.

Supplementary Information for

# Steady-State Free Precession NMR in the Presence of Heteronuclear Couplings and Decoupling: More Than Meets the Eye

Sundaresan Jayanthi, Zuzana Osifová, Mark Shif, Adonis Lupulescu, and Lucio Frydman[*]

Department of Chemical and Biological Physics, Weizmann Institute of Science, Rehovot, Israel

**Section S1: Calculation of steady-state solutions in Liouville space**

While the majority of SSFP experiments are treated in the literature in terms of magnetization vectors, the presence of J-couplings makes it natural to calculate the steady states utilizing density matrices. The homogeneous master equation[1-3] describing the evolution of the density operator is

$$\frac{d\rho}{dt} = -[i\mathcal{H} + (\Gamma + \Theta)]\rho = -\mathcal{L}\rho \qquad [S1.1]$$

The Liouvillian $\mathcal{L}$ is given by

$$\mathcal{L} = i\mathcal{H} + (\Gamma + \Theta) \qquad [S1.2]$$

where $\mathcal{H}$ represent the Hamiltonian super-matrix and $\Gamma$ is the relaxation super-matrix. The additional term $\Theta$ makes the homogeneous equation above equivalent to the inhomogeneous master equation. For a time-independent Hamiltonian the solution of Eq. [S1.1] is

$$\rho(t_2) = \exp[-\mathcal{L}(t_2 - t_1)]\rho(t_1) \qquad [S1.3]$$

In order to obtain the state of the system at any time, the exponential operator (the propagator) has to be determined. Evaluation of the propagator is always possible numerically. On the other hand, analytical expressions for the propagator are in many cases hard to obtain and/or complicated to interpret. In the present study we employed the Matlab® function 'expm' for obtaining the propagator.

Equation [3] in the main text establishing the steady state,

$$U_{pulse}U_{free}\rho_{SS} = \rho_{SS}, \qquad [S1.4]$$

can be written as

$$(U_{pulse}U_{free} - \mathbf{1})\rho_{SS} = 0 \qquad [S1.5]$$

where $\mathbf{1}$ is the unit matrix. The solution $\rho_{SS}$ is thus an eigenstate of $U_{pulse}U_{free} - \mathbf{1}$ with eigenvalue 0, hence a vector belonging to the null space. Based on this we employed Matlab's

'null' routine in order to determine $\rho_{SS}$ followed by successive computer- or 'human-' based simplifications.

The operator basis used for constructing the Liouvillian is,

$$I_Z, I_X, I_Y, S_Z, S_X, S_Y, I_Z S_Z, I_X S_Z, I_Y S_Z, I_Z S_X, I_Z S_Y, I_X S_X + I_Y S_Y \ (ZQ_X),$$

$$I_Y S_X - I_X S_Y \ (ZQ_Y), I_X S_X - I_Y S_Y \ (DQ_X), \ I_Y S_X + I_X S_Y \ (DQ_Y),$$

The Liouvillian during TR, assuming $\omega_I = 0$ and no irradiation on $I$, is

$$\begin{bmatrix}
0 & 0 & 0 & 0 & 0 & 0 & 0 & 0 & 0 & 0 & 0 & 0 & 0 & 0 & 0 & 0 \\
-4R_1 & R_1 & 0 & 0 & 0 & 0 & 0 & 0 & 0 & 0 & 0 & 0 & 0 & 0 & 0 & 0 \\
0 & 0 & R_2 & 0 & 0 & 0 & 0 & 0 & 0 & \omega_J & 0 & 0 & 0 & 0 & 0 & 0 \\
0 & 0 & 0 & R_2 & 0 & 0 & 0 & 0 & -\omega_J & 0 & 0 & 0 & 0 & 0 & 0 & 0 \\
-1R_1 & 0 & 0 & 0 & R_1 & 0 & 0 & 0 & 0 & 0 & 0 & 0 & 0 & 0 & 0 & 0 \\
0 & 0 & 0 & 0 & 0 & R_2 & \omega_S & 0 & 0 & 0 & 0 & \omega_J & 0 & 0 & 0 & 0 \\
0 & 0 & 0 & 0 & 0 & \omega_S & R_2 & 0 & 0 & 0 & -\omega_J & 0 & 0 & 0 & 0 & 0 \\
0 & 0 & 0 & 0 & 0 & 0 & 0 & R_2 & 0 & 0 & 0 & 0 & 0 & 0 & 0 & 0 \\
0 & 0 & 0 & \omega_J/4 & 0 & 0 & 0 & 0 & R_2 & 0 & 0 & 0 & 0 & 0 & 0 & 0 \\
0 & 0 & -\omega_J/4 & 0 & 0 & 0 & 0 & 0 & 0 & R_2 & 0 & 0 & 0 & 0 & 0 & 0 \\
0 & 0 & 0 & 0 & 0 & 0 & \omega_J/4 & 0 & 0 & 0 & R_2 & \omega_S & 0 & 0 & 0 & 0 \\
0 & 0 & 0 & 0 & 0 & -\omega_J/4 & 0 & 0 & 0 & 0 & -\omega_S & R_2 & 0 & 0 & 0 & 0 \\
0 & 0 & 0 & 0 & 0 & 0 & 0 & 0 & 0 & 0 & 0 & 0 & R_2 & -\omega_S & 0 & 0 \\
0 & 0 & 0 & 0 & 0 & 0 & 0 & 0 & 0 & 0 & 0 & 0 & \omega_S & R_2 & 0 & 0 \\
0 & 0 & 0 & 0 & 0 & 0 & 0 & 0 & 0 & 0 & 0 & 0 & 0 & 0 & R_2 & \omega_S \\
0 & 0 & 0 & 0 & 0 & 0 & 0 & 0 & 0 & 0 & 0 & 0 & 0 & 0 & -\omega_S & R_2
\end{bmatrix} \quad [S1.6]$$

The ratio of 4 between the correction elements (first column) accounts approximately for the ratio of the $^1H$ and $^{13}C$ gyromagnetic ratios. As mentioned in the main text, the transverse and longitudinal relaxation rates in Eq. [S1.6] wee assumed to be equal for I and S: $R_{1I} = R_{1S} = R_1, R_{2I} = R_{2S} = R_2$. The relaxation rates of other terms are assumed to be equal to $R_2$, and there were no additional cross-relaxation terms assumed in Eq. [S1.6]. While these approximations are certainly not strictly justified, they enabled us to find a 'readable' solution of Eq. [S1.4] and this solution is capable of describing the effects mentioned in the main text.

In addition to this free-evolution Liouvillian, the Liouvillian during pulses neglected any relaxation and any internal interaction, corresponding to the delta-pulse approximation. It only contained off-diagonal terms proportional to the RF amplitude $\omega_{1S}$, mixing transverse and longitudinal magnetization, zero-quantum and antiphase terms, double-quantum and antiphase terms, $I_Z S_Z$ and anti-phase terms.

For the CW decoupling case, suitable off-diagonal terms incorporating the $\omega_{1I}$ RF irradiation were added to Eq. [S1.6].

## Section 2: Numerical simulations of SSFP on J-coupled C-H systems, and their dependence on relaxation rates

Numerical SSFP Spinach simulations for a J-coupled C-H system were performed to better understand the behaviour of the SSFP magnetization and its dependence on proton and carbon relaxation rates. The simulations assumed the $'r_1, r_2'$ relaxation flag according to which all terms in the density operator relax with $R_{1H}, R_{2H}, R_{1C}, R_{2C}$, or with different sums of them, depending on the type of coherence.[4,5] For example, a double-quantum coherence would relax with $R_{2H} + R_{2C}$ etc. Figure S1 illustrates some of the outcomes of such simulations, showing the increased cross-talk between multiplet components as $R_{1H}$ increases.

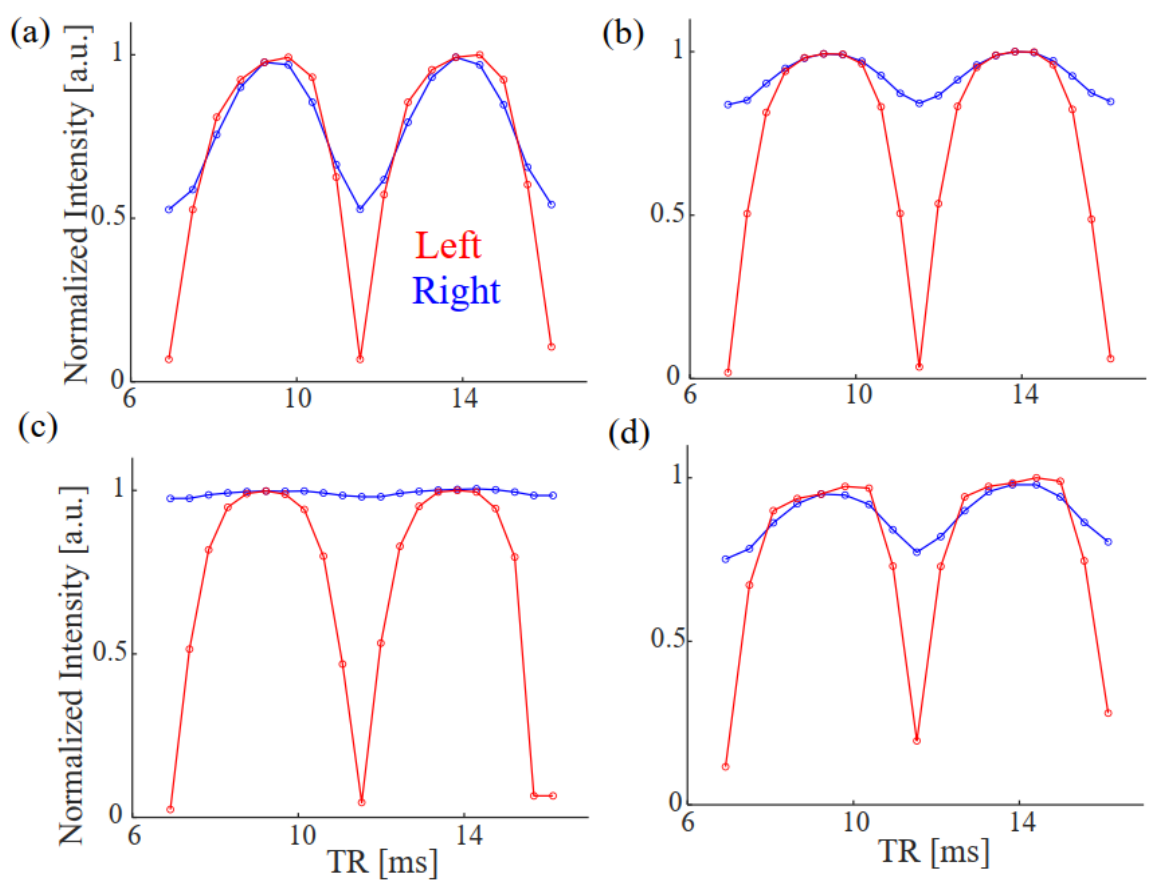

**Figure S1**. Intensity variations predicted by Spinach for the $^{13}C$ steady state magnetization as a function of $TR$ for different values of proton and carbon relaxation rates. (a) $R_{1H} = 10\ s^{-1}$; (b) $R_{1H} = 1\ s^{-1}$ (c) $R_{1H} = 0.1\ s^{-1}$ (d) $R_{1C} = R_{2C} = 10\ s^{-1}; R_{1H} = R_{2H} = 10\ s^{-1}$. In (a-c) the values $R_{1C} = 1s^{-1}; R_{2C} = 2\ s^{-1}; R_{2H} = 10\ s^{-1}$ are kept the same.

**Section 3: Experimental - NMR pulse sequences, sample preparation, relaxation rates upon addition of Cr(acac)$_3$**

NMR experiments were performed on a Bruker 600 MHz spectrometer equipped with a TCI Prodigy® probe ($^1$H at 600.4 MHz and $^{13}$C at 151.0 MHz), using an AVIII HD console and operated by Bruker's TopSpin 3.6.5 software. The SSFP sequences used were written based on the loop of equidistant pulses of a flip angle α with alternating RF phases, followed by a single pulse of flip angle α and acquisition (Fig. S2). The acquisition time, $T_{Acq}$, was set to 1.183 seconds, while the delay between pulses TR was kept in the milliseconds range. This approach –including the initial achievement of steady-states followed by acquisitions with long sampling times $T_{Acq}$, was adopted to achieve good resolution and accurate intensities for the components in the $^{13}$C doublet. The pulse sequences used are schematically described in Figure S2(a) and S2(b) respectively. The conventional decoupling schemes implemented in the Bruker TopSpin software, continuous wave (CW) and WALTZ-16. In the scheme below the phase of consecutive pulses is incremented by 180° in order to maximize signal for large flip angles α ($\alpha > 40°$). For smaller flip angles (considerably) smaller increments are needed.

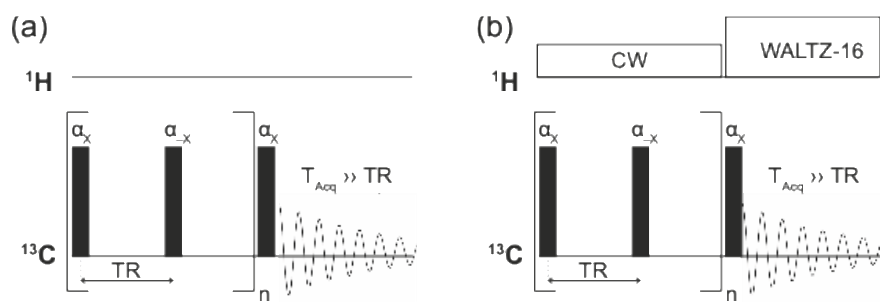

**Figure S2.** (a) Pulse sequence used in this study for monitoring $^{13}$C's SSFP behavior in the presence of J-couplings. (b) Idem but used for exploring the effects of CW decoupling during SSFP. For all experiments α = 40° (≈6 μs); for (b) a decoupling power of 0.4W was used for the WALTZ-16. *n*, the number of loops in the sequence, was varied from 262 to 1316 in order to keep the *n·TR* product constant and equal to ≈10 s before the FID acquisition.

Undoped chloroform solutions were prepared by mixing 250 μL of chloroform and 250 μL of DMSO-*d*$_6$. These chemicals were purchased from Sigma/Aldrich and Eurisotop, respectively. Samples containing Cr(acac)$_3$ were prepared by diluting a suitable 10 mM Cr(acac)$_3$ stock solution in 500 μL of a mixture of DMSO-*d*$_6$ and CHCl$_3$ (vol. 1:1). Table S1 show the obtained $^1$H relaxation times for the tested mixtures; the sample without Cr(acac)$_3$ was bubbled with nitrogen gas prior to measurement in order to exclude the effect of dissolved oxygen.

**Table S1.** Longitudinal proton relaxation times $T_1$ of the CHCl$_3$ solutions here studied at variable Cr(acac)$_3$ concentration. Subscripts A and B correspond to $^{13}$C satellites positioned at higher and lower frequencies of the $^1$H-$^{12}$C centerband (listed under $T_1$), respectively.

| Cr(acac)$_3$ [mM] | $T_1$ [s] | $T_{1,A}$ [s] | $T_{1,B}$ [s] |
|---|---|---|---|
| 0.00 | 1.376 ± 0.028 | 1.580 ± 0.032 | 1.379 ± 0.028 |
| 0.25 | 0.470 ± 0.009 | 0.485 ± 0.010 | 0.484 ± 0.010 |
| 0.50 | 0.432 ± 0.009 | 0.419 ± 0.008 | 0.421 ± 0.008 |
| 0.75 | 0.357 ± 0.007 | 0.361 ± 0.007 | 0.376 ± 0.008 |
| 1.00 | 0.279 ± 0.006 | 0.281 ± 0.006 | 0.283 ± 0.006 |

**Section 4: $^{13}$C SSFP NMR under CW decoupling: Theoretical and experimental validation for various $k$ "SSFP resonance" conditions and $\nu_e$ decoupling fields**

Figure S3 investigates the validity of Eq.'s [13,14] presented in the main text for other values of $k/\nu_e$ and $(2k+1)/2\nu_e$. It is seen that, although Eq.'s [13,14] are not exact, they nevertheless represent a fair approximation for the range of repetition rates which are relevant in SSFP experiments. In general, one can merge the two aforementioned conditions as one: $TR = m/2\nu_e$, where $m$ is an integer. Figure S4 show variation of experimental SSFP signal for different $m$ values and several $\nu_e$.

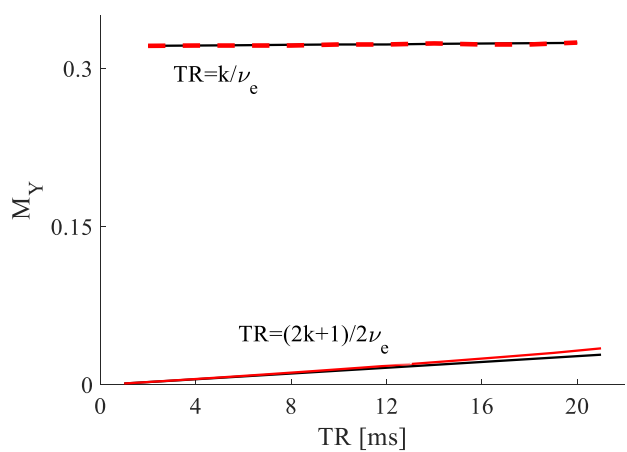

**Figure S3.** Variation SSFP ($M_Y$) with respect to TR for a $\nu_e$ of 511.5 Hz. Eq. [13,14] (black) in the main text versus brute-force numerical simulations (red). Other parameters used were $R_1 = R_2 = 1s^{-1}$ and $\alpha = 40^o$.

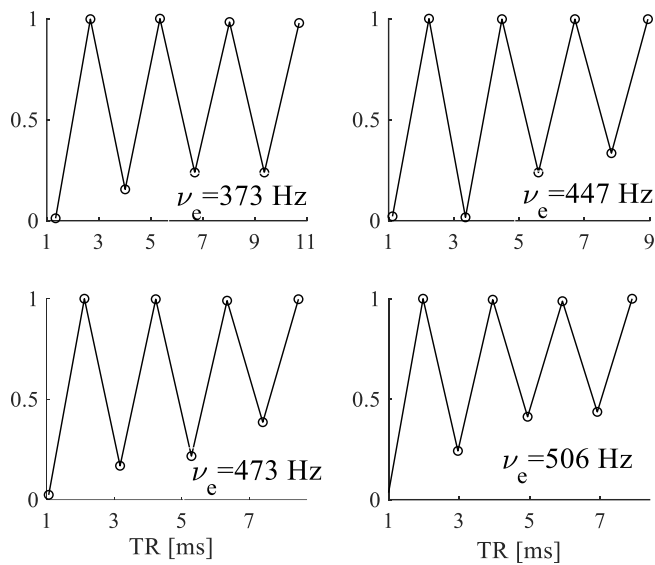

**Figure S4.** $^{13}$C SSFP signals arising from CHCl$_3$ under CW decoupling as function of TR and different RF decoupling fields. For each $\nu_e$, data were collected for $TR = m/2\nu_e$, where m is an integer (m=1,2, 3…). These data again show a very good agreement with Eqs. {13} and [14] of the main text.

**Section 5: On the depth of saturation dips versus flip angle and versus RF inhomogeneity in SSFP NMR under CW decoupling.**

Figure S5 shows the experimental dependence of SSFP dips on flip angle under CW decoupling, showing how the resonance losses increase with increasing flip angle –as adumbrated by Eq. [14] in the main text.

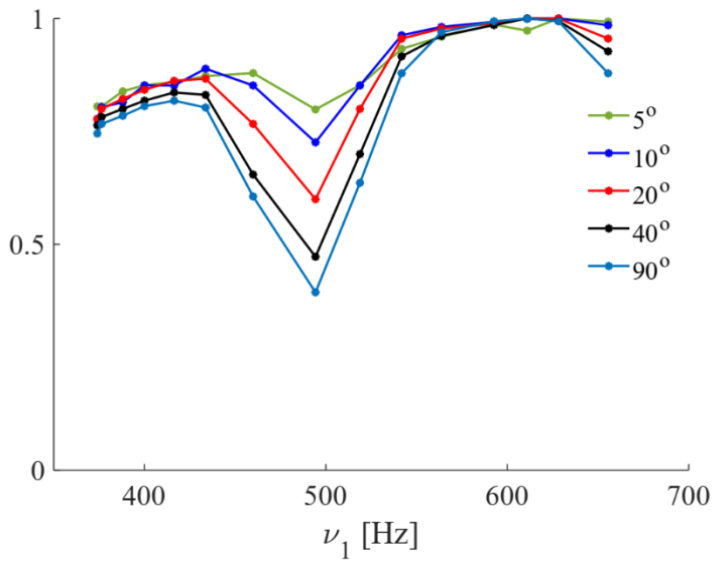

**Figure S5.** $^{13}$C SSFP signal dips of CHCl$_3$ in d$_6$-DMSO, as function of CW RF amplitude, for different flip angles. All curves were collected for a repetition time $TR$ of 5 ms and normalized to their respective maximum values.

The behaviour of the SSFP signals in the presence of CW decoupling, was further investigated with numerical simulations performed using the Spinach program. To this effect we assumed a J-coupled $^1$H-$^{13}$C spin system, undergoing SSFP on $^{13}$C and CW decoupling on the $^1$H. The SSFP signals arising after 2000 pulses arising from these simulations, are shown in Figure S6 plotted as a function of the $\nu_1$ decoupling field. These simulations were extended to incorporate CW RF inhomogeneities, by repeating calculations over 10 different $\nu_1$ values subtending a Gaussian distribution. We see that for all the cases partial saturation of the $^{13}$C SSFP signal occurs close to the $\nu^e = (2k + 1)/2TR$ (here for $k = 2$ and 3) predicted in the main text. Notice as well how RF inhomogeneity reduces the depth and increases the width of these SSFP saturation dips; best experimental fit (main text, Figure 7) was obtained with about 10% RF field inhomogeneity.

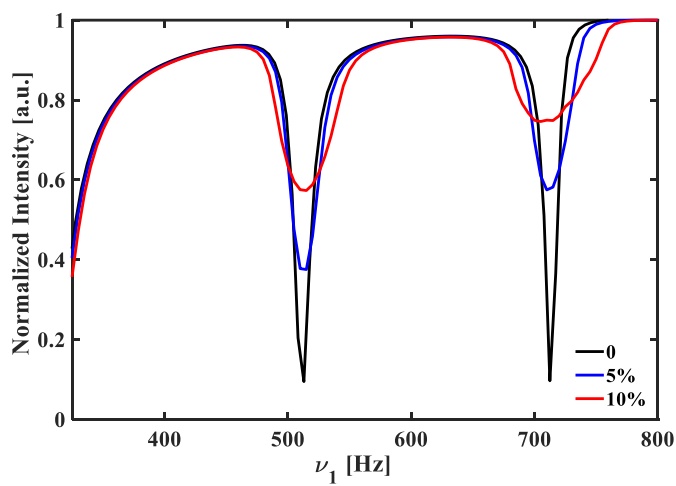

**Figure S6.** $^{13}$C SSFP signals calculated as described in the text, as a function of $\nu_1$. Black, blue and red curves assumed RF inhomogeneity spreads of 0, 5% and 10% respectively. Other variables used in simulations are $J_{IS} = 216\ Hz$, $TR = 5\ ms$, $\alpha = 40^o$, $R_1 = 2\ s^{-1}$; $R_2 = 4\ s^{-1}$.

**Section 6: On the creation of zero- and double-quantum heteronuclear states in $^1$H-decoupled SSFP $^{13}$C NMR**

Figure S7 shows the SSFP dynamics (states immediately after each pulse) of longitudinal and transverse $^{13}$C-magnetizations obtained by numerical calculations for (a) the $TR = 1/\nu^e$ and (b) the $TR = 1/2\nu^e$ conditions. Similarly, Figure S8 displays the build-up of anti-phase, zero-quantum, and double-quantum coherences arising during the SSFP experiment. As discussed in the main text, condition (b) leads to drastic reduction of magnetization. On the other hand, for *both* conditions, the zero-quantum and double-quantum coherences reach the same steady-state values, although in a very different manner. A proton antiphase coherence is also created with equal amplitudes for both conditions. Notice that carbon anti-phase coherences are absent.

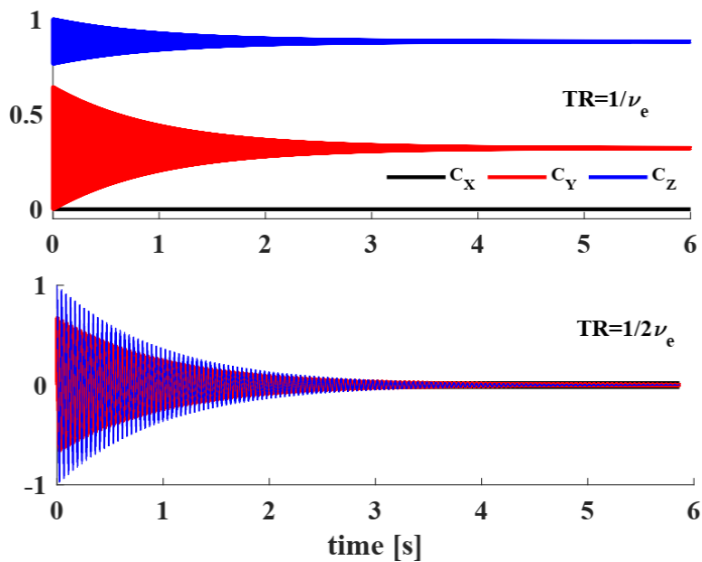

**Figure S7.** Build-up of magnetization for $TR = 1/\nu^e$ (top) and $TR = 1/2\nu^e$ (bottom). A flip angle α = 40° and on-resonance $^{13}$C pulses with x/-x phase alternations were assumed; the decoupling sideband frequency was $\nu^e = 511.5$ Hz. Other parameter used in the simulation are $J_{IS} = 216$ Hz, and $R_1 = R_2 = 1\ s^{-1}$.

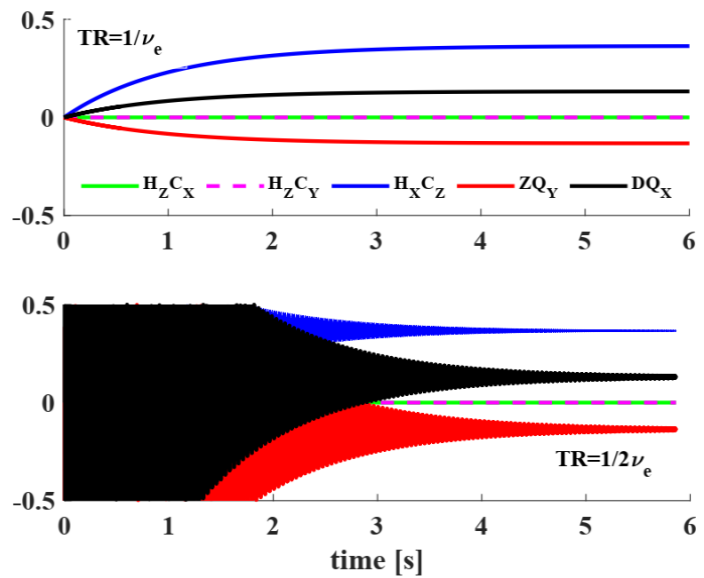

**Figure S8.** Build-up of anti-phase, zero-quantum, and double-quantum coherences for $TR = 1/\nu^e$ (top) and $TR = 1/2\nu^e$ (bottom). A flip angle of 40° was assumed and the decoupling sideband frequency is $\nu^e = 511.5$ Hz. Other parameter used in these simulation were as in Figure S7.